\documentclass[twocolumn, tighten]{aastex63}

\newcommand{\tninety}{$\tau_{90}$}
\newcommand{\tfifty}{$\tau_{50}$}

\submitjournal{APJL}

\shorttitle{Formation Histories of M31 Satellites}
\shortauthors{Weisz et al.}

\graphicspath{{./}{figures/}}

\begin{document}

\title{Comparing the Quenching Times of Faint M31 and Milky Way Satellite Galaxies}

\correspondingauthor{Daniel R. Weisz}
\email{dan.weisz@berkeley.edu}

\author{Daniel R. Weisz}
\affiliation{Department of Astronomy, University of California, Berkeley, CA 94720 USA}

\author{Nicolas F. Martin}
\affiliation{Universit\'e de Strasbourg, CNRS,  Observatoire astronomique de Strasbourg, UMR 7550, F-67000 Strasbourg France}
\affiliation{Max-Planck-Institut f\"ur Astronomie, K\"onigstuhl 17, D-69117 Heidelberg Germany}

\author{Andrew E. Dolphin}
\affiliation{Raytheon, 1151 E. Hermans Road, Tucson, AZ 85756 USA}
\affiliation{Steward Observatory, University of Arizona, 933 North Cherry Avenue, Tucson, AZ 85721-0065 USA}

\author{Saundra M. Albers}
\affiliation{Department of Astronomy, University of California, Berkeley, CA 94720 USA}

\author{Michelle L. M. Collins}
\affiliation{Department of Physics University of Surrey, Guildford, GU2 7XH, Surrey UK}

\author{Annette M. N. Ferguson}
\affiliation{Institute for Astronomy, University of Edinburgh  Royal Observatory, Blackford Hill, Edinburgh EH9 3HJ UK}

\author{Geraint F. Lewis}
\affiliation{Sydney Institute for Astronomy, School of Physics, A28, The University of Sydney, NSW 2006 Australia}

\author{Dougal Mackey}
\affiliation{Research School of Astronomy and Astrophysics, Australian
National University, Canberra, ACT 2611 Australia}

\author{Alan McConnachie}
\affiliation{National Research Council, Herzberg Institute of Astrophysics, 5071 West Saanich Road, Victoria BC V9E 2E7 Canada}

\author{R. Michael Rich}
\affiliation{Department of Physics and Astronomy, University of California, Los Angeles, PAB 430 Portola Plaza, Los Angeles, CA 90095 USA}

\author{Evan D. Skillman}
\affiliation{Minnesota Institute for Astrophysics, University of Minnesota, Minneapolis, MN 55441 USA}

\begin{abstract}
We present the star formation histories (SFHs) of 20 faint M31 satellites ($-12 \lesssim M_V \lesssim -6$) that were measured by modeling sub-horizontal branch (HB) depth color-magnitude diagrams constructed from Hubble Space Telescope (HST) imaging.  Reinforcing previous results, we find that virtually all galaxies quenched between $3$ and $9$~Gyr ago, independent of luminosity, with a notable concentration $3-6$~Gyr ago. This is in contrast to the Milky Way (MW) satellites, which are generally either faint with ancient quenching times or luminous with recent ($<$3~Gyr) quenching times.  We suggest that systematic differences in the quenching times of M31 and MW satellites may be a reflection of the varying accretion histories of M31 and the MW.  This result implies that the formation histories of low-mass satellites may not be broadly representative of low-mass galaxies in general. Among the M31 satellite population we identify two distinct groups based on their SFHs: one with exponentially declining SFHs ($\tau \sim$ 2 Gyr) and one with rising SFHs with abrupt quenching.  We speculate how these two groups could be related to scenarios for a recent major merger involving M31.  The Cycle 27 HST Treasury survey of M31 satellites will provide well-constrained ancient SFHs to go along with the quenching times we measure here.  The discovery and characterization of M31 satellites with $M_V \gtrsim -6$ would help quantify the relative contributions of reionization and environment to quenching of the lowest-mass satellites.
\end{abstract}


\section{Introduction} \label{sec:intro}

Milky Way (MW) satellite galaxies have long anchored our understanding of low-mass mass galaxy formation.  Their number counts, spatial distributions, and structural properties are used to constrain dark matter cosmology on small scales \citep[e.g.,][]{bullock2017, simon2019}.  Their star formation histories (SFHs) and chemical content provide insight into cosmic reionization and the baryonic processes that uniquely shape low-mass galaxy evolution \citep[e.g.,][]{kirby2011, brown2014, weisz2014a}. More recently, their orbital histories, as measured by the Hubble Space Telescope (HST) and Gaia, reveal the complex effects of central galaxies on the evolution of low-mass satellites \citep[e.g.,][]{sohn2012, fritz2018, simon2018}.

At the same time, there is growing evidence that the MW satellites may not be representative of low-mass satellites in general. Compared to the MW, satellite systems throughout the local Universe show varying luminosity functions, stellar populations, quenching properties, and spatial configurations, often in excess of cosmic variance \citep[e.g.,][]{mcconnachie2006, brasseur2011, Tollerud:2011wd, geha2017, muller2018, pawlowski2018, smercina2018, pawlowski2019}. Even in our nearest neighbor, M31, there are hints that the internal \citep[e.g., kinematics, stellar content;][]{dacosta1996, collins2010, collins2014, martin2017} and global \citep[e.g., ``plane of satellites'';][]{ibata2013, pawlowski2018} properties of M31 and MW satellites are different. Thus, it is unclear whether the fundamental insights established in MW satellites are applicable to all low-mass systems or stem from the specific accretion history of the MW.

In this Letter, we present the first uniform SFH measurements of many faint M31 satellites.  We use sub-horizontal branch (HB) HST imaging, which has been presented in two previous papers in this series \citep{martin2017, weisz2019a}, to measure their SFHs, and we compare our measurements to literature SFHs of MW satellites.   Relative to the MW satellites, the SFHs we measure for M31 satellites have coarser age resolution, because the M31 CMDs do not reach the oldest main sequence turn off (MSTO).  We will acquire oldest MSTO photometry for all known M31 satellites during HST Cycle 27 as part of HST-GO-15902 (PI D.~Weisz), which will strengthen the results we preview in this Letter.

\begin{figure*}[ht!]
\begin{center}
\includegraphics[scale=0.65]{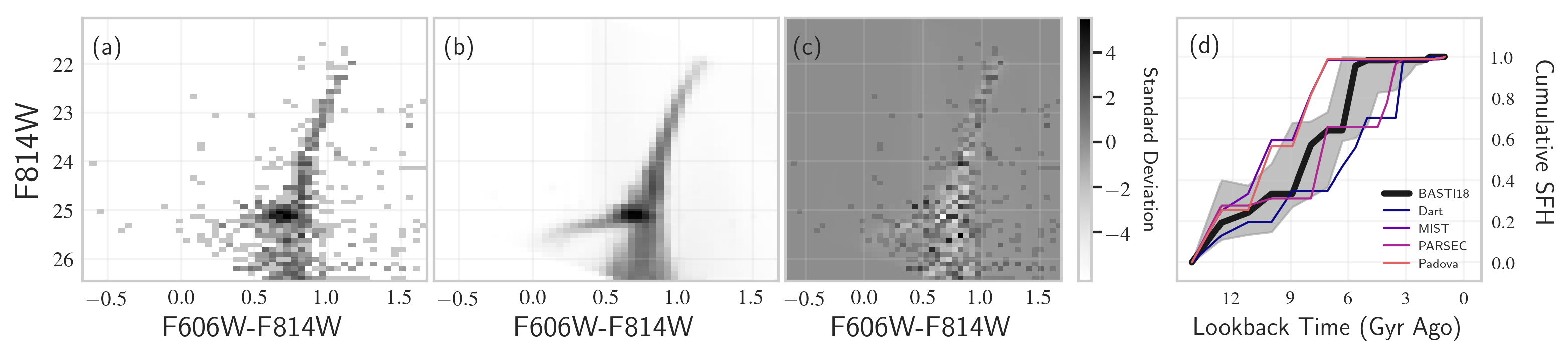}
\caption{An example BASTI-based SFH measurement using And~XXI.  Panel (a) shows the observed Hess diagram. Panel (b) shows the best fit model Hess diagram for the BaSTI stellar evolution models. Panel (c) shows the residual significance diagram, i.e., (data-model)/model, in units of standard deviation.  The lack of systematic structures in Panel (c), i.e., large contiguous regions of black or white points, indicates a good fit.  Panel (d) shows the cumulative SFH, i.e., the fraction of stars formed before a given epoch.  The solid black line is the best fit SFH for the BaSTI models and the grey shaded envelope are the total (random plus systematic) uncertainties for the 68\% confidence interval.  The thin colored lines are the best fit SFHs for different stellar evolution models.  Overall, the shape of the SFH is similar between the models, but the ages of SFH features can shift by $\sim 2$~Gyr owing to differences in the underlying stellar physics.  The grey shaded region captures the 1-$\sigma$ scatter in the different SFHs reasonably well.}
\label{fig:4panel}
\end{center}
\end{figure*}

\section{The Data} \label{sec:data}

The observations and data reduction used in this program are described in detail in \citet{martin2017}. Here, we provide a brief summary.  

Through HST-GO-13699 (PI N.~Martin), we observed 16 faint M31 satellites with the Advanced Camera for Surveys (ACS) 

that had no previous HST imaging. Each galaxy was observed for a single orbit with equal integration times in the F606W and F814W filters.  We added archival F606W and F814W ACS data of comparable depth for And~XVIII (HST-SNAP-13442; PI B.~Tully).  We also added archival observations of And~XI, And~XII, and And~XIII (HST-GO-11084; PI D.~Zucker) which were taken in F606W and F814W with the Wide Field Planetary Camera 2 (WFPC2).
  
In total, our sample has 20 systems with $-12 \lesssim M_V \lesssim -6$, including 7 ultra-faint dwarf galaxies (UFDs, $M_{V} > -7.7$; \citealt{simon2019}).

For each galaxy, we performed point spread function photometry with \texttt{DOLPHOT}, a widely-used package for reducing observations of resolved stellar populations with HST-specific modules \citep{dolphin2000b}. We adopted the \texttt{DOLPHOT} parameters recommended in \citet{williams2014}. The raw photometric catalogs were culled to include only good stars as described in \citet{martin2017}.  We ran $\sim 10^5$ artificial star tests (ASTs) for each galaxy to quantify completeness and photometric uncertainties.  The 50\% completeness limit for a typical galaxy in our sample is $F606W \sim 27.1$ and $F814W \sim 26.2$.

Figure \ref{fig:4panel} illustrates the quality of our data by showing a Hess diagram, i.e., a binned CMD, for And~XXI ($M_{V} = -9.2$). The CMD shows a clear red giant branch (RGB), red clump (RC), and a predominantly red HB, as described in \citet{martin2017}.  The faint limit of the CMD is set to the 50\% completeness limit of $F814W =26.3$, which is $\sim$1.5 magnitudes fainter than the HB. 

\section{Methodology} \label{sec:methods}

We measure the SFHs of the 20 galaxies in our sample using \texttt{MATCH} \citep{dolphin2002}, a software package that forward models the CMD of a galaxy in order to measure its SFH as described in \citet{weisz2014a}.  Here, we provide a brief summary of the pertinent details.

For this analysis, we adopt a Kroupa IMF \citep{kroupa2001}, a binary fraction of 0.35, HB-based distances for the ACS data \citep{weisz2019a} and self-consistent TRGB-based distances for the WFPC2 data \citep{weisz2014a}, and foreground extinction values from \citet{schlafly2011}.  

We fit each entire CMD with five different stellar evolution libraries: Dartmouth \citep{dotter2008}, Padova \citep{girardi2010}, PARSEC \citep{bressan2012}, MIST \citep{choi2016}, and BaSTI \citep{hidalgo2018}. We find that for all CMDs analyzed in this paper the BaSTI 2018 models provide the best overall fits in terms of visual inspection of the residuals and through comparison of likelihood ratios between models.  Thus, we adopt the BaSTI models for this paper.  

We adopt a metallicity grid that ranges from $-2.3 \le$ [M/H] $\le -0.5$ with a resolution of 0.1 dex and an age grid that ranges from $9.00 \le \log(t/{\rm yr}) \le 10.15$ with a resolution of $\log(t/{\rm yr}) = 0.05$ dex. We find that including ages younger than $\log(t/{\rm yr}) =9.0$ did not change the SFHs (as these galaxies have no young populations) but increased the computational time. Thus, we simply exclude ages with $\log(t/{\rm yr}) < 9.0$ from our CMD modeling.

Finally, given that our CMDs do not reach the oldest MSTO, we follow \citet{weisz2011a} and
\citet{weisz2014a} and adopt a prior on the age-metallicity relationship that requires the metallicity to increase monotonically with time, with a modest dispersion allowed at each age.  This choice helps to mitigate some of the age-metallicity degeneracy on the RGB and RC.   We compute random (which account for the finite number of stars on the a CMD) and systematic uncertainties (which are a proxy for uncertainties in the physics of the underlying stellar models) as described in \citet{weisz2014a}.  Finally, as detailed in \S 3.6 and Figure 6 of \citet{weisz2014a}, SFHs measured from CMDs that include the HB but do not include the oldest MSTO have larger systematic uncertainties, but are consistent with SFHs measured from oldest MSTO-depth CMDs.

Figure \ref{fig:4panel} illustrates the CMD modeling process.  Panel (a) shows the observed Hess diagram of And~XXI and panel (b) shows the best fit model for the BaSTI stellar library.  A visual comparison of the model and data indicates good overall agreement.  Panel (c) is a residual significance diagram, i.e., (model-data)/model, which quantifies the level of (dis)agreement.  The majority of populated pixels are consistent within $\sim$ 1-$\sigma$, while only a handful of pixels are highly discrepant (i.e., $>$ 3-$\sigma$). There are no signs of poorly fit regions of the CMD (e.g., large swaths of only black or white pixels), which indicates that the model is a good fit to the data.

\begin{figure*}[ht!]
\begin{center}
\includegraphics[scale=0.7]{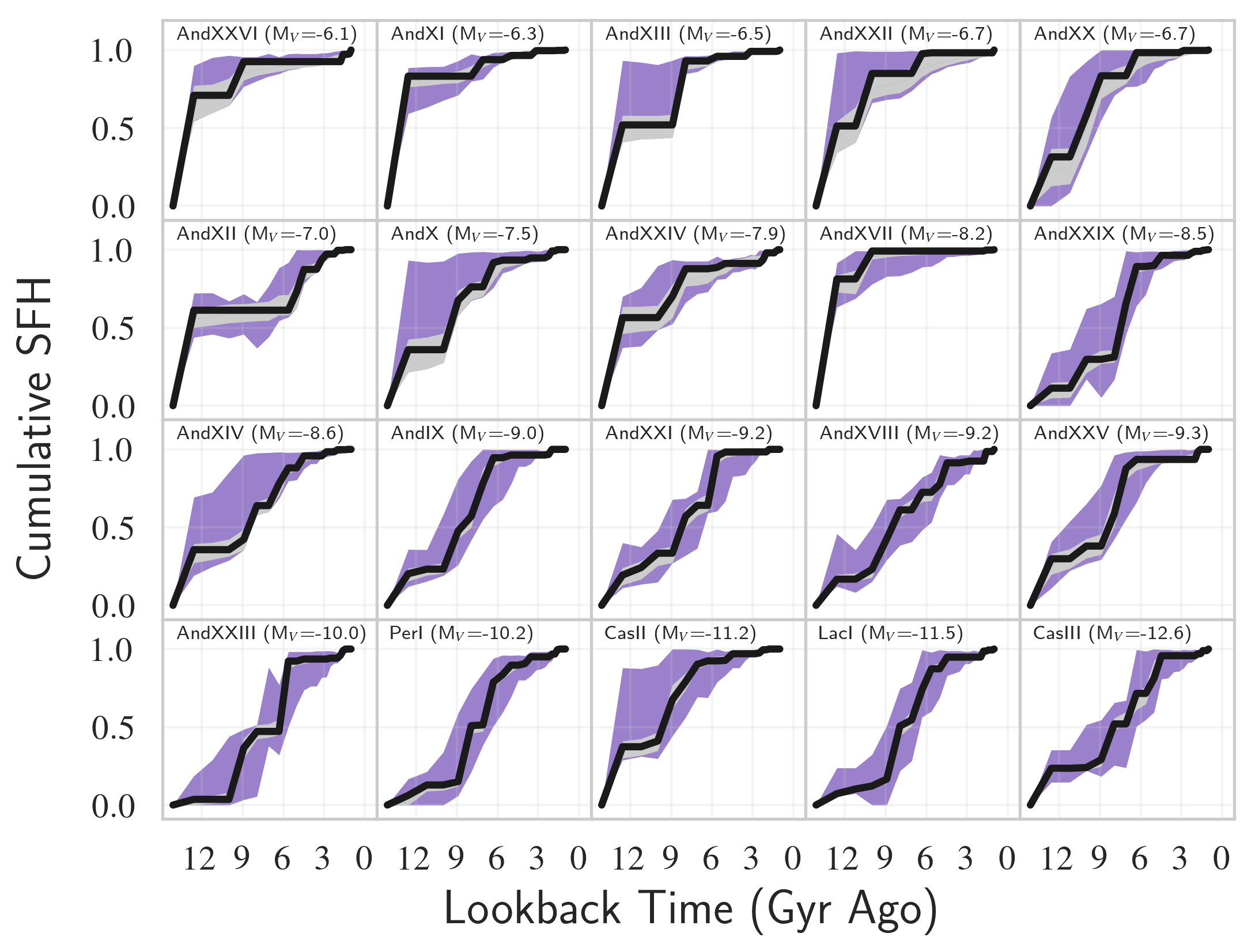}
\caption{The cumulative SFHs of 20 faint M31 satellites ordered by increasing luminosity.  The black solid line is the best fit BaSTI SFH.  The grey and purple shaded envelopes reflect the 68\% confidence intervals for the random and total uncertainties, respectively. Fainter M31 satellites generally form the bulk of their stellar mass at earlier time compared to the brighter systems. All galaxies appear to have quenching times between 3 and 9 Gyr ago.    \label{fig:sfhs}}
\end{center}
\end{figure*}

Panel (d) shows the cumulative SFH of And~XXI, i.e., the fraction of total stellar mass formed prior to a given epoch.  The solid black line is the best fit BaSTI solution and the grey shaded band reflects the 68\% confidence interval of the total (i.e., random plus systematic) uncertainties. The random uncertainties are negligibly small compared to the systematic uncertainties.  

The thin colored lines in panel (d) are the best fit SFHs from the other four stellar libraries.  These SFHs are similar in shape to the BaSTI solution, though particlar features can be shifted by up to $\sim 2$ Gyr, which is due to differences in the underlying stellar physics \citep[e.g.,][]{Gallart:2005qy}.  As intended, the scatter in the SFHs from different models is well-captured by the grey-shaded error envelope. 

The small amount of apparent star formation $\sim$ 1-3 Gyr ago is likely due to a handful of blue stragglers which occupy similar portions of the CMD as younger main sequence stars \citep[e.g.,][]{monelli2012}.

\begin{figure}[ht!]
\begin{center}
\includegraphics[scale=0.7]{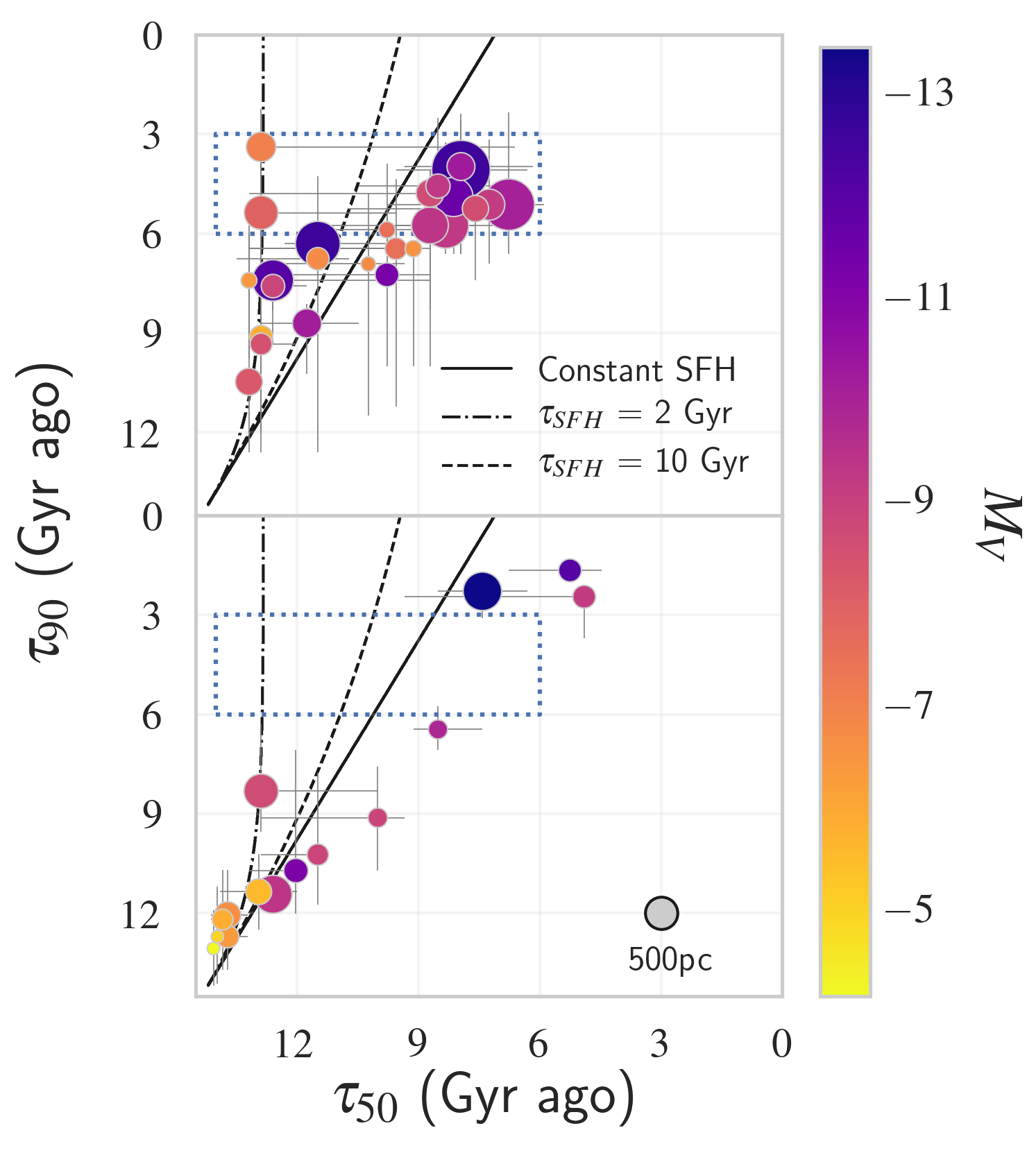}
\caption{The lookback time (Gyr ago) at which 50\% of the stellar mass formed (\tfifty) versus the time at which 90\% of the stellar mass formed (\tninety), i.e., the quenching time.  The top panel includes the 20 M31 satellites from this paper and 6 from \citet{skillman2017}.  Points are color-coded by luminosity and their relative sizes reflect their half-light radii.  The grey point indicates a size of 500~pc.  The black lines illustrate a constant and exponentially declining SFHs.  The bottom panels shows results from literature SFHs of MW satellites.  The area enclosed by the blue dotted line contains half the M31 sample, but no MW satellites. The smaller uncertainties for the \citet{skillman2017} M31 dSphs are indicative of what can be expected from the forthcoming cycle 27 observations. \label{fig:t50_t90}}
\end{center}
\end{figure}

\section{Results and Discussion} \label{sec:discussion}

Figure \ref{fig:sfhs} shows the cumulative SFHs of 20 faint M31 satellites plotted in order of increasing luminosity from upper left to lower right.  The solid black lines are the best fit SFHs, while the grey and purple shaded envelopes reflect the 68\% confidence intervals for the random and total (random plus systematic) uncertainties, respectively.  These values are tabulated in Table \ref{tab:sfhs}.

This figure reveals both a diversity of SFHs among the M31 faint satellite population and some broad trends.  For example,  galaxies with $M_V \gtrsim -8.5$ tend to have formed $\gtrsim 50$\% of their stars prior to $\sim 10-12$~Gyr ago, compared to 6-9~Gyr ago for more luminous systems.  Interestingly, galaxies such as And~XXIX, Per~I, and Lac~I appear to have formed $\lesssim$ 10\% of their stellar mass prior to $\sim 10$~Gyr ago, which is unusually low when put into context with our knowledge of LG dwarf galaxy SFHs \citep[e.g.,][]{weisz2014a, gallart2015}. 

Another interesting feature is the quenching times.  That is, very few systems have either very early ($>$10-12 Gyr ago) or very late ($\lesssim 3$~Gyr ago) quenching times.  Instead, the vast majority of the systems stopped forming stars $\sim 3-6$ Gyr ago, almost independent of luminosity.  

The top panel of Figure \ref{fig:t50_t90} consolidates the SFHs of the 20 faint M31 satellites from this paper and the 6 systems from \citet{skillman2017} into a more digestible form. Here, we plot the quenching time\footnote{Following the literature \citep[e.g.,][]{weisz2015a, skillman2017}, we adopt the time at which 90\% of the total stellar mass formed as a proxy for the quenching time to avoid ambiguity due to blue stragglers.} (\tninety) versus the time at which 50\% of the total stellar mass formed (\tfifty).  Points are color-coded by luminosity, point sizes are proportional to half-light radius, and the error bars reflect the total (random plus systematic) uncertainties.  To guide the eye, we overplot black lines that illustrate cases of a constant (solid) and exponentially declining SFHs ($\tau_{SFH}=10$~Gyr, dashed lined; $\tau_{SFH}=2$~Gyr, dot-dashed lined).  

This plot shows several interesting trends.  First, although there are several predominantly ancient galaxies (i.e., \tfifty $>12$~Gyr), there are very few systems with \tninety $>12$~Gyr. Instead, the predominantly ancient systems have a range of \tninety\ values that extend from 3 to 10 Gyr ago.  This is particularly interesting in the context of reionization, in which the prevailing view is that the lowest-mass galaxies have star formation shutdown by reionization in the very early Universe \citep[e.g.,][]{ bullock2000, benson2002, ricotti2005}.  However, we caution against over-interpretation of this finding, as (i) we lack a complete census of UFDs around M31 and (ii) current SFHs are uncertain at the oldest ages.

Second, there are no galaxies that quenched within the last 3 Gyr, i.e., \tninety\ $<3$~Gyr ago.  Instead, most galaxies have quenching times concentrated at intermediate ages, with a notable grouping at \tninety$\sim$ 3-6~Gyr ago.  

Third, there appears to be some degree of synchronicity in the quenching of some systems.  Most notably, the over-density of M31 satellites at \tfifty $\sim 6-9$~Gyr ago also all seem to have \tninety\ $\sim 3-6$~Gyr ago, with no clear trends in galaxy size or luminosity, in agreement with the results of the representative sample from \citet{skillman2017}.

Fourth, compared to the over-plotted fiducial SFH models (i.e., constant SFH, exponentially declining) there are two distinct groups. Eight galaxies fall tightly on the $\tau_{SFH}=2$~Gyr track while 15 are to the right of the constant SFH track (i.e., they have rising SFHs). Only three galaxies exist between these groups.

The age information provided by our SFHs may help model the complex formation and accretion history of the M31 and its satellites.  The M31 halo hosts rich stellar substructures (e.g., streams, over-densities) that suggest an active history of mergers in M31 (see \citealt{mcconnachie2018} and references therein) and its stellar halo and outer disk have large populations of intermediate age stars, as revealed by oldest MSTO-depth CMD analysis \citep[e.g.,][]{brown2006a, bernard2015b}.  Several models have posited a major merger between M31 and (what would have been) the third largest member of the LG $\sim$2-4 Gyr ago \citep[e.g.,][]{hammer2018, dsouza2018}.  These models can qualitatively explain some observed features of M31, such as a global burst of star formation 2-4 Gyr ago \citep[e.g.,][]{bernard2015b, williams2017} and the metal-rich inner halo.  For example, \citet{dsouza2018} hypothesize such an interaction between M31 and M32p (the putative progenitor of M32) could explain the metal-rich component of M31's halo and the unusually compact nature of M32.  This model implies that M32p had a $M_{\star} \sim 2.5\times 10^{10}$ $M_{\odot}$ prior to its interaction with M31, making it the third largest member of the LG just a few Gyr ago.  One implication of this scenario may be that the large number of satellites with \tninety $\sim$ 3-6~Gyr ago may have been environmentally quenched during the merger. A second speculative angle is that the dichotomy of SFHs in the top panel of Figure \ref{fig:t50_t90}, i.e., rising SFHs vs exponentially declining, may be due to the presence of two difference satellite populations, i.e., one set from M31, the other from M32p.  Though speculative, we use these examples to illustrate the potential of our data for deciphering the formation history of M31's halo and emphasize that more rigorous analysis is clearly warranted.

We also consider the relationship between our SFHs and sub-structures in the M31 system, e.g., the plane of satellites from \citet{ibata2013}.  We find no clear evidence for a correlation with membership in structures  identified in \citet{ibata2013} and \citet{santossantos2019}.  However, given the large uncertainties and unclear theoretical expectations between sub-structures and SFHs, the lack of a clear correlation is challenging to interpret.

Figure \ref{fig:t50_t90} also summarizes differences in the formation histories of M31 and MW satellites.  In the bottom panel, we plot \tninety\ \emph{vs.} \tfifty\ for the MW satellites using literature SFHs \citep[e.g.,][]{brown2014, weisz2015a} and global properties \citet[e.g., luminosity, size;][]{mcconnachie2012}.  

It is striking that the M31 and MW satellite populations do not share many similar trends.  The M31 satellites fill out intermediate values of \tfifty\ and \tninety, i.e., $6 \lesssim$ \tfifty\ $\lesssim 12$~Gyr ago and $3 \lesssim$ \tninety\ $\lesssim 6$~Gyr ago (the dotted blue box in Figure \ref{fig:t50_t90}, whereas there are essentially no MW satellites in that range.  In terms of quenching, the MW satellites Fornax, Carina, and Leo~I (galaxies located in the upper right region of the lower panel) all ceased star formation within the most recent $\sim 1-3$~Gyr, whereas none of the M31 satellites did. The faintest MW satellites ($M_V \gtrsim -7$) all quenched $\gtrsim 12$~Gyr ago, presumably due to reionization \cite[e.g.,][]{brown2014}, but some of the comparably faint M31 satellites appear to have more extended SFHs. 

This may indicate that the evolution of the satellites are coupled to the accretion history of the host galaxy.  By extension, it may be that the MW satellites do not cover the full range of intrinsic formation histories of low-mass galaxies.

\begin{deluxetable}{lcccc}
\tabletypesize{\footnotesize}
\tablecaption{Summary statistics for SFHs of faint M31 satellites. Values of \tfifty\ and \tninety\ are from the best fit SFHs measured in this paper.  Error bars are the 68\% confidence intervals for the total uncertainties (i.e., random plus systematic).  Values for the galaxies below the horizontal lines are taken directly from \citet{skillman2017}. }
\label{tab:sfhs}
\tablehead{
\colhead{Name} & \colhead{$M_V$} & 
\colhead{$r_h$} & \colhead{\tfifty} & 
\colhead{\tninety} \\ 
\colhead{} & \colhead{(mag)} & 
\colhead{(pc)} & \colhead{(Gyr ago)} & 
\colhead{(Gyr ago)} \\
\colhead{(1)} & \colhead{(2)} & 
\colhead{(3)} & \colhead{(4)} & 
\colhead{(5)} 
} 
\startdata
Cas~III & $-12.6$ & 1640 & 7.9$_{-1.6}^{+1.6}$ & 4.1$_{-1.5}^{+2.5}$\\
Lac~I & $-11.5$ & 967 & 8.1$_{-1.7}^{+0.8}$ & 4.9$_{-1.7}^{+1.7}$\\
Cas~II & $-11.2$ & 275 & 9.8$_{-1.1}^{+3.4}$ & 7.2$_{-3.4}^{+2.8}$\\
Per~I & $-10.2$ & 384 & 7.9$_{-1.8}^{+1.4}$ & 4.0$_{-1.6}^{+2.6}$\\
And~XXIII & $-10.0$ & $1277$ & 6.8$_{-0.9}^{+1.9}$ & 5.1$_{-2.8}^{+1.5}$\\
And~XXV & $-9.3$ & 679 & 8.7$_{-1.1}^{+2.8}$ & 5.8$_{-1.3}^{+2.6}$\\
And~XXI & $-9.2$ & 1033 & 8.3$_{-1.9}^{+1.2}$ & 5.8$_{-2.5}^{+0.9}$\\
And~XVIII & $-9.2$ & 262 & 8.5$_{-1.6}^{+2.0}$ & 4.6$_{-2.1}^{+1.7}$\\
And~IX & $-9.0$ & 444 & 7.2$_{-0.3}^{+2.5}$ & 5.1$_{-2.0}^{+1.8}$\\
And~XIV & $-8.6$ & 379 & 8.7$_{-0.4}^{+4.5}$ & 4.8$_{-0.7}^{+5.2}$\\
And~XXIX & $-8.5$ & 397 & 7.6$_{-0.7}^{+3.1}$ & 5.2$_{-1.2}^{+2.2}$\\
And~XVII & $-8.2$ & $339$ & 13.2$_{-0.3}^{+0.0}$ & 10.5$_{-5.0}^{+2.1}$\\
And~XXIV & $-7.9$ & $579$ & 12.9$_{-3.3}^{+0.3}$ & 5.4$_{-3.1}^{+4.4}$\\
And~X & $-7.5$ & 239 & 9.5$_{-0.2}^{+3.6}$ & 6.5$_{-2.1}^{+4.8}$\\
And~XII & $-7.0$ & 420 & 12.9$_{-6.3}^{+0.3}$ & 3.4$_{-0.2}^{+2.6}$\\
And~XXII & $-6.7$ & 253 & 11.5$_{-0.8}^{+2.0}$ & 6.8$_{-2.5}^{+5.8}$\\
And~XX & $-6.7$ & 110 & 10.2$_{-0.9}^{+2.6}$ & 6.9$_{-2.1}^{+4.6}$\\
And~XIII & $-6.5$ & $130$ & 9.1$_{-0.4}^{+4.1}$ & 6.5$_{-0.1}^{+3.5}$\\
And~XI & $-6.3$ & 120 & 13.2$_{-0.3}^{+0.0}$ & 7.4$_{-1.4}^{+2.4}$\\
And~XXVI & $-6.1$ & 228 & 12.9$_{-0.3}^{+0.3}$ & 9.1$_{-6.0}^{+2.9}$\\
\hline 
\hline
And~II & $-12.6$ & 965 & 11.5$_{-2.1}^{+0.8}$ & 6.3$_{-0.6}^{+0.5}$\\
And~I & $-12.0$ & 815 & 12.6$_{-3.9}^{+0.3}$ & 7.4$_{-0.7}^{+0.9}$ \\
And~III & $-10.1$ & 405 & 11.7$_{-1.3}^{+1.1}$ & 8.7$_{-0.6}^{+1.5}$\\
And~XXVIII & $-8.8$ & 265 & 12.6$_{-0.8}^{+0.3}$ & 7.6$_{-0.3}^{+1.7}$\\
And~XV & $-8.4$ & 230 & 12.9$_{-0.9}^{+0.3}$ & 9.3$_{-0.8}^{+3.3}$\\
And~XVI & $-7.5$ & 130 & 9.8$_{-1.1}^{+1.4}$ & 5.9$_{-0.6}^{+0.4}$
\enddata
\end{deluxetable}

There are several caveats with the present analysis. First, while SFHs from MW satellites are all measured from CMDs that reach the oldest MSTO, our new M31 data are much shallower.  Consequently, we are left with large uncertainties that may hide various trends in the data. Moreover, our SFHs are based primarily on the HB morphology, which is not as well an understood phase of stellar evolution as the MSTO \citep[e.g.,][]{Gallart:2005qy}. However, we note that comparisons of SFHs measured from different depths (i.e., HB vs. oldest MSTO) generally show good agreement \citep[e.g.,][]{weisz2014a} as previously described. We urge appropriate caution against over-interpreting this generation of M31 SFHs, particularly at ancient epochs.  
Second, there are various selection effects that we have not explicitly considered.  One is the size of the HST field of view relative to the size of a galaxy.  In some cases, this can lead to $\sim 1$~Gyr biases in the measured SFH relative to the true global SFH \citep[e.g.,][]{graus2019b}.  Another is the lack of many known UFDs in the M31 ecosystem.  Detecting faint systems is quite challenging at the distance of M31, given the paucity of bright stars and the high level of contamination.  

Despite these challenges, we are optimistic about prospects of placing the M31 satellites onto equal observational footing with their MW counterparts.  The 244 orbit cycle 27 HST Treasury program (HST-GO-15902; PI D.~Weisz) will obtain MSTO depth imaging across the entire M31 satellite system, which will significantly reduce the uncertainties on the SFHs and establish a first epoch for proper motion measurements. 

In our view, among the most important next steps for M31 satellites is to identify UFDs around M31. If UFDs around M31 are found to have substantially extended SFHs, then our picture of how reionization affects low-mass galaxy formation, halo occupation, etc.\ may fundamentally change.  Finding and characterizing UFDs around M31 requires dedicated imaging and spectroscopic efforts, as well as the power of HST and/or JWST for measuring their star formation and orbital histories.  

\section*{Acknowledgements}

Support for HST program GO-13699 was provided by NASA through a grant from the Space Telescope Science Institute, which is operated by the Association of Universities for Research in Astronomy, Incorporated, under NASA contract NAS5-26555. These observations are associated with program HST-SNAP-13442 and HST-GO-13699. DRW acknowledges support from an Alfred P. Sloan Fellowship and an Alexander von Humboldt Fellowship. SMA is supported by the National Science Foundation Graduate Research Fellowship under Grant DGE 1752814. This research has made use of the NASA/IPAC Extragalactic Database (NED) which is operated by the Jet Propulsion Laboratory, California Institute of Technology, under contract with the National Aeronautics and Space Administration.

\bibliography{m31_sfhs_astroph}{}

\begin{thebibliography}{}
\expandafter\ifx\csname natexlab\endcsname\relax\def\natexlab#1{#1}\fi
\providecommand{\url}[1]{\href{#1}{#1}}
\providecommand{\dodoi}[1]{doi:~\href{http://doi.org/#1}{\nolinkurl{#1}}}
\providecommand{\doeprint}[1]{\href{http://ascl.net/#1}{\nolinkurl{http://ascl.net/#1}}}
\providecommand{\doarXiv}[1]{\href{https://arxiv.org/abs/#1}{\nolinkurl{https://arxiv.org/abs/#1}}}

\bibitem[{{Benson} {et~al.}(2002){Benson}, {Frenk}, {Lacey}, {Baugh}, \&
  {Cole}}]{benson2002}
{Benson}, A.~J., {Frenk}, C.~S., {Lacey}, C.~G., {Baugh}, C.~M., \& {Cole}, S.
  2002, \mnras, 333, 177, \dodoi{10.1046/j.1365-8711.2002.05388.x}

\bibitem[{{Bernard} {et~al.}(2015){Bernard}, {Ferguson}, {Chapman}, {Ibata},
  {Irwin}, {Lewis}, \& {McConnachie}}]{bernard2015b}
{Bernard}, E.~J., {Ferguson}, A.~M.~N., {Chapman}, S.~C., {et~al.} 2015,
  \mnras, 453, L113, \dodoi{10.1093/mnrasl/slv116}

\bibitem[{{Brasseur} {et~al.}(2011){Brasseur}, {Martin}, {Macci{\`o}}, {Rix},
  \& {Kang}}]{brasseur2011}
{Brasseur}, C.~M., {Martin}, N.~F., {Macci{\`o}}, A.~V., {Rix}, H.-W., \&
  {Kang}, X. 2011, \apj, 743, 179, \dodoi{10.1088/0004-637X/743/2/179}

\bibitem[{{Bressan} {et~al.}(2012){Bressan}, {Marigo}, {Girardi}, {Salasnich},
  {Dal Cero}, {Rubele}, \& {Nanni}}]{bressan2012}
{Bressan}, A., {Marigo}, P., {Girardi}, L., {et~al.} 2012, \mnras, 427, 127,
  \dodoi{10.1111/j.1365-2966.2012.21948.x}

\bibitem[{{Brown} {et~al.}(2006){Brown}, {Smith}, {Guhathakurta}, {Rich},
  {Ferguson}, {Renzini}, {Sweigart}, \& {Kimble}}]{brown2006a}
{Brown}, T.~M., {Smith}, E., {Guhathakurta}, P., {et~al.} 2006, \apjl, 636,
  L89, \dodoi{10.1086/500089}

\bibitem[{{Brown} {et~al.}(2014){Brown}, {Tumlinson}, {Geha}, {Simon},
  {Vargas}, {VandenBerg}, {Kirby}, {Kalirai}, {Avila}, {Gennaro}, {Ferguson},
  {Mu{\~n}oz}, {Guhathakurta}, \& {Renzini}}]{brown2014}
{Brown}, T.~M., {Tumlinson}, J., {Geha}, M., {et~al.} 2014, \apj, 796, 91,
  \dodoi{10.1088/0004-637X/796/2/91}

\bibitem[{{Bullock} \& {Boylan-Kolchin}(2017)}]{bullock2017}
{Bullock}, J.~S., \& {Boylan-Kolchin}, M. 2017, \araa, 55, 343,
  \dodoi{10.1146/annurev-astro-091916-055313}

\bibitem[{{Bullock} {et~al.}(2000){Bullock}, {Kravtsov}, \&
  {Weinberg}}]{bullock2000}
{Bullock}, J.~S., {Kravtsov}, A.~V., \& {Weinberg}, D.~H. 2000, \apj, 539, 517,
  \dodoi{10.1086/309279}

\bibitem[{{Choi} {et~al.}(2016){Choi}, {Dotter}, {Conroy}, {Cantiello},
  {Paxton}, \& {Johnson}}]{choi2016}
{Choi}, J., {Dotter}, A., {Conroy}, C., {et~al.} 2016, \apj, 823, 102,
  \dodoi{10.3847/0004-637X/823/2/102}

\bibitem[{{Collins} {et~al.}(2010){Collins}, {Chapman}, {Irwin}, {Martin},
  {Ibata}, {Zucker}, {Blain}, {Ferguson}, {Lewis}, {McConnachie}, \&
  {Pe{\~n}arrubia}}]{collins2010}
{Collins}, M.~L.~M., {Chapman}, S.~C., {Irwin}, M.~J., {et~al.} 2010, \mnras,
  407, 2411, \dodoi{10.1111/j.1365-2966.2010.17069.x}

\bibitem[{{Collins} {et~al.}(2014){Collins}, {Chapman}, {Rich}, {Ibata},
  {Martin}, {Irwin}, {Bate}, {Lewis}, {Pe{\~n}arrubia}, {Arimoto}, {Casey},
  {Ferguson}, {Koch}, {McConnachie}, \& {Tanvir}}]{collins2014}
{Collins}, M.~L.~M., {Chapman}, S.~C., {Rich}, R.~M., {et~al.} 2014, \apj, 783,
  7, \dodoi{10.1088/0004-637X/783/1/7}

\bibitem[{{Da Costa} {et~al.}(1996){Da Costa}, {Armandroff}, {Caldwell}, \&
  {Seitzer}}]{dacosta1996}
{Da Costa}, G.~S., {Armandroff}, T.~E., {Caldwell}, N., \& {Seitzer}, P. 1996,
  \aj, 112, 2576, \dodoi{10.1086/118204}

\bibitem[{{Dolphin}(2000)}]{dolphin2000b}
{Dolphin}, A.~E. 2000, \pasp, 112, 1383, \dodoi{10.1086/316630}

\bibitem[{{Dolphin}(2002)}]{dolphin2002}
---. 2002, MNRAS, 332, 91, \dodoi{10.1046/j.1365-8711.2002.05271.x}

\bibitem[{{Dotter} {et~al.}(2008){Dotter}, {Chaboyer}, {Jevremovi{\'c}},
  {Kostov}, {Baron}, \& {Ferguson}}]{dotter2008}
{Dotter}, A., {Chaboyer}, B., {Jevremovi{\'c}}, D., {et~al.} 2008, \apjs, 178,
  89, \dodoi{10.1086/589654}

\bibitem[{{D'Souza} \& {Bell}(2018)}]{dsouza2018}
{D'Souza}, R., \& {Bell}, E.~F. 2018, Nature Astronomy, 2, 737,
  \dodoi{10.1038/s41550-018-0533-x}

\bibitem[{{Fritz} {et~al.}(2018){Fritz}, {Battaglia}, {Pawlowski},
  {Kallivayalil}, {van der Marel}, {Sohn}, {Brook}, \& {Besla}}]{fritz2018}
{Fritz}, T.~K., {Battaglia}, G., {Pawlowski}, M.~S., {et~al.} 2018, ArXiv
  e-prints, arXiv:1805.00908.
\newblock \doarXiv{1805.00908}

\bibitem[{{Gallart} {et~al.}(2005){Gallart}, {Zoccali}, \&
  {Aparicio}}]{Gallart:2005qy}
{Gallart}, C., {Zoccali}, M., \& {Aparicio}, A. 2005, \araa, 43, 387,
  \dodoi{10.1146/annurev.astro.43.072103.150608}

\bibitem[{{Gallart} {et~al.}(2015){Gallart}, {Monelli}, {Mayer}, {Aparicio},
  {Battaglia}, {Bernard}, {Cassisi}, {Cole}, {Dolphin}, {Drozdovsky},
  {Hidalgo}, {Navarro}, {Salvadori}, {Skillman}, {Stetson}, \&
  {Weisz}}]{gallart2015}
{Gallart}, C., {Monelli}, M., {Mayer}, L., {et~al.} 2015, \apjl, 811, L18,
  \dodoi{10.1088/2041-8205/811/2/L18}

\bibitem[{{Geha} {et~al.}(2017){Geha}, {Wechsler}, {Mao}, {Tollerud}, {Weiner},
  {Bernstein}, {Hoyle}, {Marchi}, {Marshall}, {Mu{\~n}oz}, \& {Lu}}]{geha2017}
{Geha}, M., {Wechsler}, R.~H., {Mao}, Y.-Y., {et~al.} 2017, \apj, 847, 4,
  \dodoi{10.3847/1538-4357/aa8626}

\bibitem[{{Girardi} {et~al.}(2010){Girardi}, {Williams}, {Gilbert},
  {Rosenfield}, {Dalcanton}, {Marigo}, {Boyer}, {Dolphin}, {Weisz},
  {Melbourne}, {Olsen}, {Seth}, \& {Skillman}}]{girardi2010}
{Girardi}, L., {Williams}, B.~F., {Gilbert}, K.~M., {et~al.} 2010, \apj, 724,
  1030, \dodoi{10.1088/0004-637X/724/2/1030}

\bibitem[{{Graus} {et~al.}(2019){Graus}, {Bullock}, {Fitts}, {Cooper},
  {Boylan-Kolchin}, {Weisz}, {Wetzel}, {Feldmann}, {Faucher-Gigu{\`e}re},
  {Quataert}, {Hopkins}, \& {Keres}}]{graus2019b}
{Graus}, A.~S., {Bullock}, J.~S., {Fitts}, A., {et~al.} 2019, arXiv e-prints,
  arXiv:1901.05487.
\newblock \doarXiv{1901.05487}

\bibitem[{{Hammer} {et~al.}(2018){Hammer}, {Yang}, {Wang}, {Ibata}, {Flores},
  \& {Puech}}]{hammer2018}
{Hammer}, F., {Yang}, Y.~B., {Wang}, J.~L., {et~al.} 2018, Monthly Notices of
  the Royal Astronomical Society, 475, 2754, \dodoi{10.1093/mnras/stx3343}

\bibitem[{{Hidalgo} {et~al.}(2018){Hidalgo}, {Pietrinferni}, {Cassisi},
  {Salaris}, {Mucciarelli}, {Savino}, {Aparicio}, {Silva Aguirre}, \&
  {Verma}}]{hidalgo2018}
{Hidalgo}, S.~L., {Pietrinferni}, A., {Cassisi}, S., {et~al.} 2018, \apj, 856,
  125, \dodoi{10.3847/1538-4357/aab158}

\bibitem[{{Ibata} {et~al.}(2013){Ibata}, {Lewis}, {Conn}, {Irwin},
  {McConnachie}, {Chapman}, {Collins}, {Fardal}, {Ferguson}, {Ibata}, {Mackey},
  {Martin}, {Navarro}, {Rich}, {Valls-Gabaud}, \& {Widrow}}]{ibata2013}
{Ibata}, R.~A., {Lewis}, G.~F., {Conn}, A.~R., {et~al.} 2013, \nat, 493, 62,
  \dodoi{10.1038/nature11717}

\bibitem[{{Kirby} {et~al.}(2011){Kirby}, {Lanfranchi}, {Simon}, {Cohen}, \&
  {Guhathakurta}}]{kirby2011}
{Kirby}, E.~N., {Lanfranchi}, G.~A., {Simon}, J.~D., {Cohen}, J.~G., \&
  {Guhathakurta}, P. 2011, \apj, 727, 78, \dodoi{10.1088/0004-637X/727/2/78}

\bibitem[{{Kroupa}(2001)}]{kroupa2001}
{Kroupa}, P. 2001, \mnras, 322, 231, \dodoi{10.1046/j.1365-8711.2001.04022.x}

\bibitem[{{Martin} {et~al.}(2017){Martin}, {Weisz}, {Albers}, {Bernard},
  {Collins}, {Dolphin}, {Ferguson}, {Ibata}, {Lewis}, {Mackey}, {McConnachie},
  {Rich}, \& {Skillman}}]{martin2017}
{Martin}, N.~F., {Weisz}, D.~R., {Albers}, S.~M., {et~al.} 2017, 1704.01586v1.
\newblock \doarXiv{1704.01586v1}

\bibitem[{{McConnachie}(2012)}]{mcconnachie2012}
{McConnachie}, A.~W. 2012, \aj, 144, 4, \dodoi{10.1088/0004-6256/144/1/4}

\bibitem[{{McConnachie} \& {Irwin}(2006)}]{mcconnachie2006}
{McConnachie}, A.~W., \& {Irwin}, M.~J. 2006, \mnras, 365, 1263,
  \dodoi{10.1111/j.1365-2966.2005.09806.x}

\bibitem[{{McConnachie} {et~al.}(2018){McConnachie}, {Ibata}, {Martin},
  {Ferguson}, {Collins}, {Gwyn}, {Irwin}, {Lewis}, {Mackey}, {Davidge},
  {Arias}, {Conn}, {C{\^o}t{\'e}}, {Crnojevic}, {Huxor}, {Penarrubia},
  {Spengler}, {Tanvir}, {Valls-Gabaud}, {Babul}, {Barmby}, {Bate}, {Bernard},
  {Chapman}, {Dotter}, {Harris}, {McMonigal}, {Navarro}, {Puzia}, {Rich},
  {Thomas}, \& {Widrow}}]{mcconnachie2018}
{McConnachie}, A.~W., {Ibata}, R., {Martin}, N., {et~al.} 2018, \apj, 868, 55,
  \dodoi{10.3847/1538-4357/aae8e7}

\bibitem[{{Monelli} {et~al.}(2012){Monelli}, {Cassisi}, {Mapelli}, {Bernard},
  {Aparicio}, {Skillman}, {Stetson}, {Gallart}, {Hidalgo}, {Mayer}, \&
  {Tolstoy}}]{monelli2012}
{Monelli}, M., {Cassisi}, S., {Mapelli}, M., {et~al.} 2012, \apj, 744, 157,
  \dodoi{10.1088/0004-637X/744/2/157}

\bibitem[{{M{\"u}ller} {et~al.}(2018){M{\"u}ller}, {Jerjen}, \&
  {Binggeli}}]{muller2018}
{M{\"u}ller}, O., {Jerjen}, H., \& {Binggeli}, B. 2018, ArXiv e-prints,
  arXiv:1802.08657.
\newblock \doarXiv{1802.08657}

\bibitem[{{Pawlowski}(2018)}]{pawlowski2018}
{Pawlowski}, M.~S. 2018, Modern Physics Letters A, 33, 1830004,
  \dodoi{10.1142/S0217732318300045}

\bibitem[{{Pawlowski} {et~al.}(2019){Pawlowski}, {Bullock}, {Kelley}, \&
  {Famaey}}]{pawlowski2019}
{Pawlowski}, M.~S., {Bullock}, J.~S., {Kelley}, T., \& {Famaey}, B. 2019, \apj,
  875, 105, \dodoi{10.3847/1538-4357/ab10e0}

\bibitem[{{Ricotti} \& {Gnedin}(2005)}]{ricotti2005}
{Ricotti}, M., \& {Gnedin}, N.~Y. 2005, \apj, 629, 259, \dodoi{10.1086/431415}

\bibitem[{{Santos-Santos} {et~al.}(2019){Santos-Santos},
  {Dom{\'\i}nguez-Tenreiro}, \& {Pawlowski}}]{santossantos2019}
{Santos-Santos}, I.~M., {Dom{\'\i}nguez-Tenreiro}, R., \& {Pawlowski}, M.~S.
  2019, arXiv e-prints, arXiv:1908.02298.
\newblock \doarXiv{1908.02298}

\bibitem[{{Schlafly} \& {Finkbeiner}(2011)}]{schlafly2011}
{Schlafly}, E.~F., \& {Finkbeiner}, D.~P. 2011, \apj, 737, 103,
  \dodoi{10.1088/0004-637X/737/2/103}

\bibitem[{{Simon}(2018)}]{simon2018}
{Simon}, J.~D. 2018, ArXiv e-prints, arXiv:1804.10230.
\newblock \doarXiv{1804.10230}

\bibitem[{{Simon}(2019)}]{simon2019}
---. 2019, arXiv e-prints, arXiv:1901.05465.
\newblock \doarXiv{1901.05465}

\bibitem[{{Skillman} {et~al.}(2017){Skillman}, {Monelli}, {Weisz}, {Hidalgo},
  {Aparicio}, {Bernard}, {Boylan-Kolchin}, {Cassisi}, {Cole}, {Dolphin},
  {Ferguson}, {Gallart}, {Irwin}, {Martin}, {Mart{\'{\i}}nez-V{\'a}zquez},
  {Mayer}, {McConnachie}, {McQuinn}, {Navarro}, \& {Stetson}}]{skillman2017}
{Skillman}, E.~D., {Monelli}, M., {Weisz}, D.~R., {et~al.} 2017, \apj, 837,
  102, \dodoi{10.3847/1538-4357/aa60c5}

\bibitem[{{Smercina} {et~al.}(2018){Smercina}, {Bell}, {Price}, {D'Souza},
  {Slater}, {Ballin}, {Monachesi}, \& {Nidever}}]{smercina2018}
{Smercina}, A., {Bell}, E.~F., {Price}, P.~A., {et~al.} 2018, ArXiv e-prints,
  arXiv:1807.03779.
\newblock \doarXiv{1807.03779}

\bibitem[{{Sohn} {et~al.}(2012){Sohn}, {Anderson}, \& {van der
  Marel}}]{sohn2012}
{Sohn}, S.~T., {Anderson}, J., \& {van der Marel}, R.~P. 2012, \apj, 753, 7,
  \dodoi{10.1088/0004-637X/753/1/7}

\bibitem[{{Tollerud} {et~al.}(2011){Tollerud}, {Bullock}, {Graves}, \&
  {Wolf}}]{Tollerud:2011wd}
{Tollerud}, E.~J., {Bullock}, J.~S., {Graves}, G.~J., \& {Wolf}, J. 2011, \apj,
  726, 108, \dodoi{10.1088/0004-637X/726/2/108}

\bibitem[{{Weisz} {et~al.}(2014){Weisz}, {Dolphin}, {Skillman}, {Holtzman},
  {Gilbert}, {Dalcanton}, \& {Williams}}]{weisz2014a}
{Weisz}, D.~R., {Dolphin}, A.~E., {Skillman}, E.~D., {et~al.} 2014, \apj, 789,
  147, \dodoi{10.1088/0004-637X/789/2/147}

\bibitem[{{Weisz} {et~al.}(2015){Weisz}, {Dolphin}, {Skillman}, {Holtzman},
  {Gilbert}, {Dalcanton}, \& {Williams}}]{weisz2015a}
---. 2015, \apj, 804, 136, \dodoi{10.1088/0004-637X/804/2/136}

\bibitem[{{Weisz} {et~al.}(2011){Weisz}, {Dalcanton}, {Williams}, {Gilbert},
  {Skillman}, {Seth}, {Dolphin}, {McQuinn}, {Gogarten}, {Holtzman}, {Rosema},
  {Cole}, {Karachentsev}, \& {Zaritsky}}]{weisz2011a}
{Weisz}, D.~R., {Dalcanton}, J.~J., {Williams}, B.~F., {et~al.} 2011, \apj,
  739, 5, \dodoi{10.1088/0004-637X/739/1/5}

\bibitem[{{Weisz} {et~al.}(2019){Weisz}, {Dolphin}, {Martin}, {Albers},
  {Collins}, {Ferguson}, {Lewis}, {Mackey}, {McConnachie}, {Rich}, \&
  {Skillman}}]{weisz2019a}
{Weisz}, D.~R., {Dolphin}, A.~E., {Martin}, N.~F., {et~al.} 2019, \mnras, 1928,
  \dodoi{10.1093/mnras/stz1984}

\bibitem[{{Williams} {et~al.}(2014){Williams}, {Lang}, {Dalcanton}, {Dolphin},
  {Weisz}, {Bell}, {Bianchi}, {Byler}, {Gilbert}, {Girardi}, {Gordon},
  {Gregersen}, {Johnson}, {Kalirai}, {Lauer}, {Monachesi}, {Rosenfield},
  {Seth}, \& {Skillman}}]{williams2014}
{Williams}, B.~F., {Lang}, D., {Dalcanton}, J.~J., {et~al.} 2014, \apjs, 215,
  9, \dodoi{10.1088/0067-0049/215/1/9}

\bibitem[{{Williams} {et~al.}(2017){Williams}, {Dolphin}, {Dalcanton}, {Weisz},
  {Bell}, {Lewis}, {Rosenfield}, {Choi}, {Skillman}, \&
  {Monachesi}}]{williams2017}
{Williams}, B.~F., {Dolphin}, A.~E., {Dalcanton}, J.~J., {et~al.} 2017, \apj,
  846, 145, \dodoi{10.3847/1538-4357/aa862a}

\end{thebibliography}
\bibliographystyle{aasjournal}

\end{document}